 \documentclass[twocolumn]{webofc}
\usepackage[varg]{txfonts}   
\usepackage{hyperref}
\usepackage{url}
\hypersetup{colorlinks=true,citecolor=blue,urlcolor=blue,linkcolor=blue}
\begin{document}
\title{High-granularity Dual-readout Calorimeter:  Evolution of a Classic Prototype}
%

\author{
\firstname{Nural} \lastname{Akchurin}\inst{}\fnsep\thanks{\email{Nural.Akchurin@ttu.edu}}\and
\firstname{James} \lastname{Cash} \and
\firstname{Jordan} \lastname{Damgov}\and
\firstname{Xander} \lastname{Delashaw}\and
\firstname{Kamal} \lastname{Lamichhane} \and
\firstname{Miles} \lastname{Harris}\and
\firstname{Mitch} \lastname{Kelley} \and
\firstname{Shuichi} \lastname{Kunori} \and
\firstname{Harold} \lastname{Mergate-Cacace} \and
\firstname{Timo} \lastname{Peltola}\and
\firstname{Odin} \lastname{Schneider} \and
\firstname{Julian} \lastname{Sewell}
}

\institute{Advanced Particle Detector Laboratory,
Department of Physics and Astronomy,
Texas Tech University, Lubbock, TX, USA
          }

\abstract{The original dual-readout calorimeter prototype (DREAM), constructed two decades ago, has proven instrumental in advancing our understanding of calorimetry.  It has facilitated a multitude of breakthroughs by leveraging signals from complementary media (Cherenkov and scintillation) to capture fluctuations in electromagnetic energy fraction within hadronic showers.  Over the years, extensive studies have shed light on the performance characteristics of this module, rendering it exceptionally well-understood.  Drawing on this wealth of experience, we have embarked on enhancing the detectors’ capabilities further by integrating fast silicon photomultipliers (SiPMs) with finer transverse segmentation, $\sim$1 cm$^2$, as well as longitudinal segmentation by timing measuring better than 10 cm.  This configuration will allow us to image hadronic showers with high granularity (HG-DREAM).  We argue that the spatial information provided by such a granular detector in a short time window ($\approx$5 ns) leads to substantial enhancement in energy resolution when advanced neural networks are employed in energy reconstruction. We briefly present the current status of work, new concepts that have been introduced to the detector, and expectations from simulations.
}
\maketitle
\section{Introduction}
\label{intro}
\begin{figure}[h]
\centering
\includegraphics[width=8cm,clip]{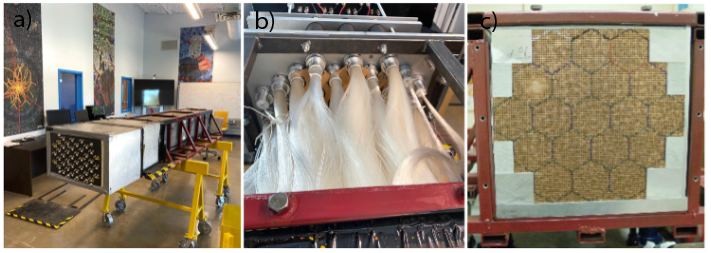}
\caption{\small The original DREAM calorimeter before reconfiguration: with the PMT readout box (a), large scintillating and clear fiber bundles behind the absorber (b), and the transverse segmentation with 19 hexagonal towers (c).  The tower size is $r\approx3.71$ cm.}
\label{fig:original_dream}       
\end{figure}

The original DREAM calorimeter was built over two decades ago and ushered in a new approach to calorimetry.  The characteristics of this prototype are well-known and documented over twenty publications.  The detector consisted of plastic scintillation and plastic/fused-silica clear fibers uniformly embedded in a 1,030-kg copper matrix consisting of 5,580 2-m long square rods (4$\times$4 mm$^2$) with a small hole in the center (dia. 2.5 mm).  The detector was originally segmented into 19 towers where each tower, hexagonal in shape, measured 3.71 cm in radius.  There was no longitudinal segmentation in depth (Figure~\ref{fig:original_dream}).  The characteristic parameters were $X_{\rm 0}=20.1$ mm, $\rho_{\rm M}=20.4$ mm, and $\lambda_{\rm I}=200$ mm.  By volume, the prototype consisted of 69.2\% Cu, 9.4\% scintillating fibers, 12.6\% clear fibers, and 8.7\% air.  The sampling frequency for scintillating fibers was 2.1\%.  The scintillating fibers, SCSF-81J, were purchased from Kuraray; clear plastic fibers, Raytela PJR-FB750, from Toray; and the fused-silica fibers, hard-polymer clad fused-silica core, from Polymicro.  The power of this calorimeter stemmed from the simultaneous measurement of the Cherenkov ($Q$) and scintillation ($S$) signals for each event and thereby accurately estimating the electromagnetic energy fraction in an event from the $Q/S$ ratio.  This approach enabled us to remove the contribution of the fluctuations in the electromagnetic energy fraction to the energy resolution event-by-event, vastly improving the hadronic energy reconstruction. 

In addition to improved energy resolution compared to traditional techniques, this approach resulted in a linear response to hadrons.  In other words, the dual-readout approach compensated ($e/h=1$) the hardronic response event-by-event.  The electromagnetic energy resolution for the scintillation part was $23.7\%/\sqrt{E}\oplus2.8\%$ and for the Cherenkov part was $37.5\%/\sqrt{E}\oplus2.6\%$ when the detector was tilted a few degrees with the respect to the beam direction in order to avoid channeling of the beam particles along the fibers~\cite{Akchurin:2005eu}.  The hadronic energy resolution for single pions for the scintillation part was  $49\%/\sqrt{E} + 7\%$ and for the Cherenkov part was $86\%/\sqrt{E}+10\%$.  After correction for $f_{\rm em}$ fluctuations using the $Q/S$ ratio, the energy resolution improved to $41\%/\sqrt{E} + 4.2\%$ and the hadronic energy response became linear within $\pm3\%$~\cite{Akchurin:2005an}. 

\section{High-granularity DREAM (HG-DREAM)}
\label{sec:HG-DREAM}

\begin{figure}[!htbp]
\centering
\includegraphics[width=8.3cm,clip]{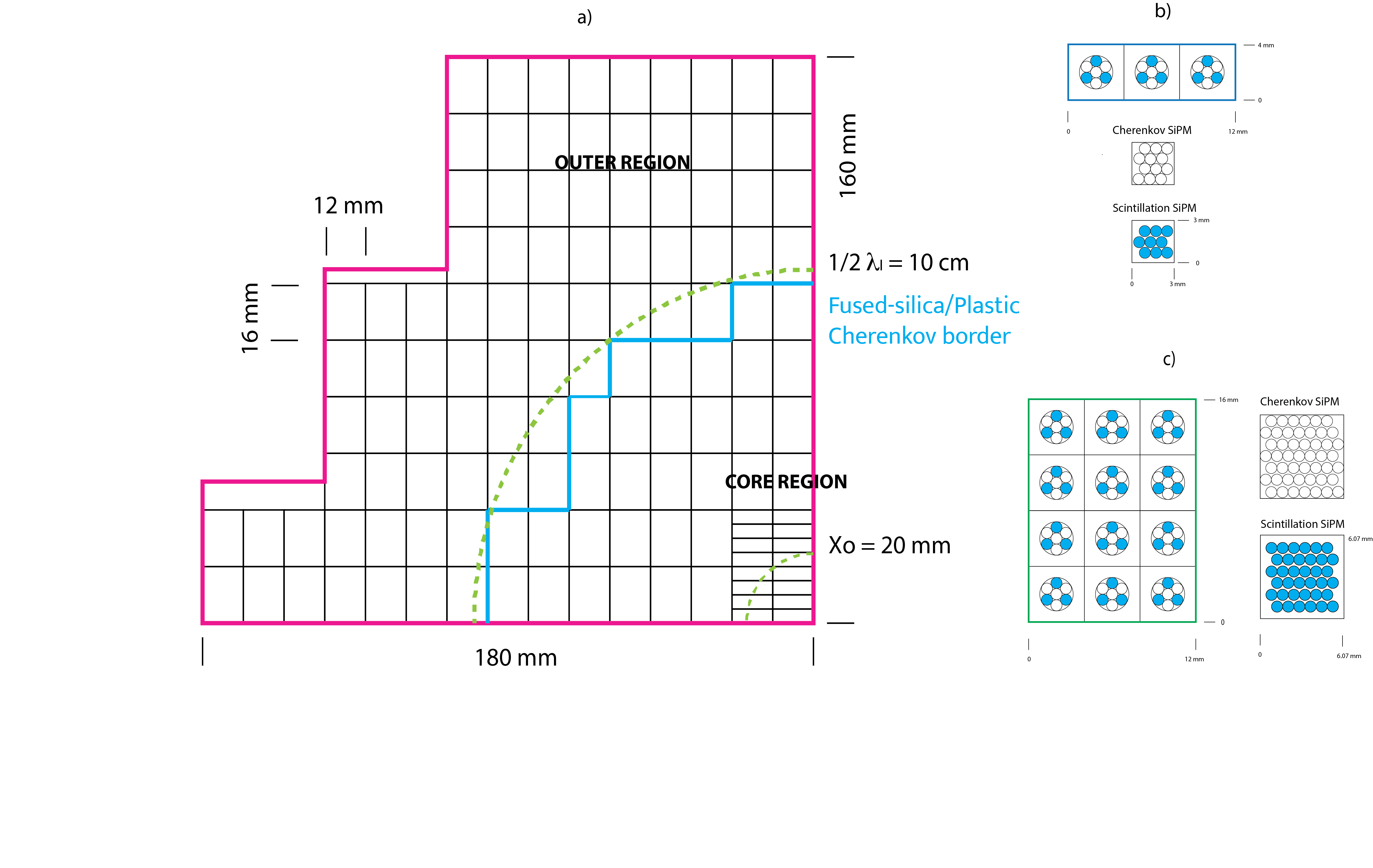}
\caption{\small HG-DREAM transverse segmentation and fiber configuration:  (a) the detector is segmented into rectangular towers (a quarter is shown); (b) the tower size in the core of the detector is relatively small, consisting of 1$\times$3 rods, where the scintillating and clear fibers are bundled separately and routed to 3$\times$3 mm$^2$ SiPMs.  The towers in the outer region are four times larger (12 copper rods in a 3$\times$4 arrangement) and the separately bundled fibers terminate at 6$\times$6 mm$^2$ SiPMs.  The plastic/fused-silica clear fibers are indicated in white, and the scintillating fibers are in blue. }
\label{fig:hg-dream_configuration}      
\end{figure}

The recent developments in SiPMs and readout electronics and the application of neural networks to image analysis open new avenues for calorimetry.  Reconfiguring the DREAM module into smaller transverse towers using SiPMs and effectively longitudinally segmenting by timing without changing the sampling fractions turns DREAM into an ``imaging'' calorimeter, or HG-DREAM.   By doing this, we expect to demonstrate that sufficiently granular longitudinal segmentation by timing is feasible.  In conjunction with the application of neural networks, this approach promises a paradigm shift beyond what’s possible with the original DREAM prototype.   Figure~\ref{fig:hg-dream_configuration}.a shows the high-granularity reconfiguration.  A tower is represented by a rectangular shape that contains 3$\times$4 square rods (1.2$\times$1.6 cm$^2$), and Cherenkov and scintillation photons are read out separately by 6$\times$6 mm$^2$ OnSemi SiPMs (Figure~\ref{fig:hg-dream_configuration}.c).  The central core of the detector is segmented more finely, $1\times3$ square rods measuring 0.4$\times$1.2 cm$^2$, and is read out by 3$\times$3 mm$^2$ OnSemi SiPMs (Figure~\ref{fig:hg-dream_configuration}.b).  The HG-DREAM is $\sim24\times$ more finely segmented compared to the original DREAM module.  In the core region, we use 128, and in the outer region we use 768 OnSemi SiPMs (Type MicroFJ with 35 $\mu$m pitch).  We take advantage of the SiPM fast outputs, with $\sim10\times$ amplification, for timing measurement (CAEN DRS V1742), and of the standard outputs for energy measurement (CAEN FERS 5200 system).  

\begin{figure}[!htbp]
\centering
\includegraphics[width=8cm,clip]{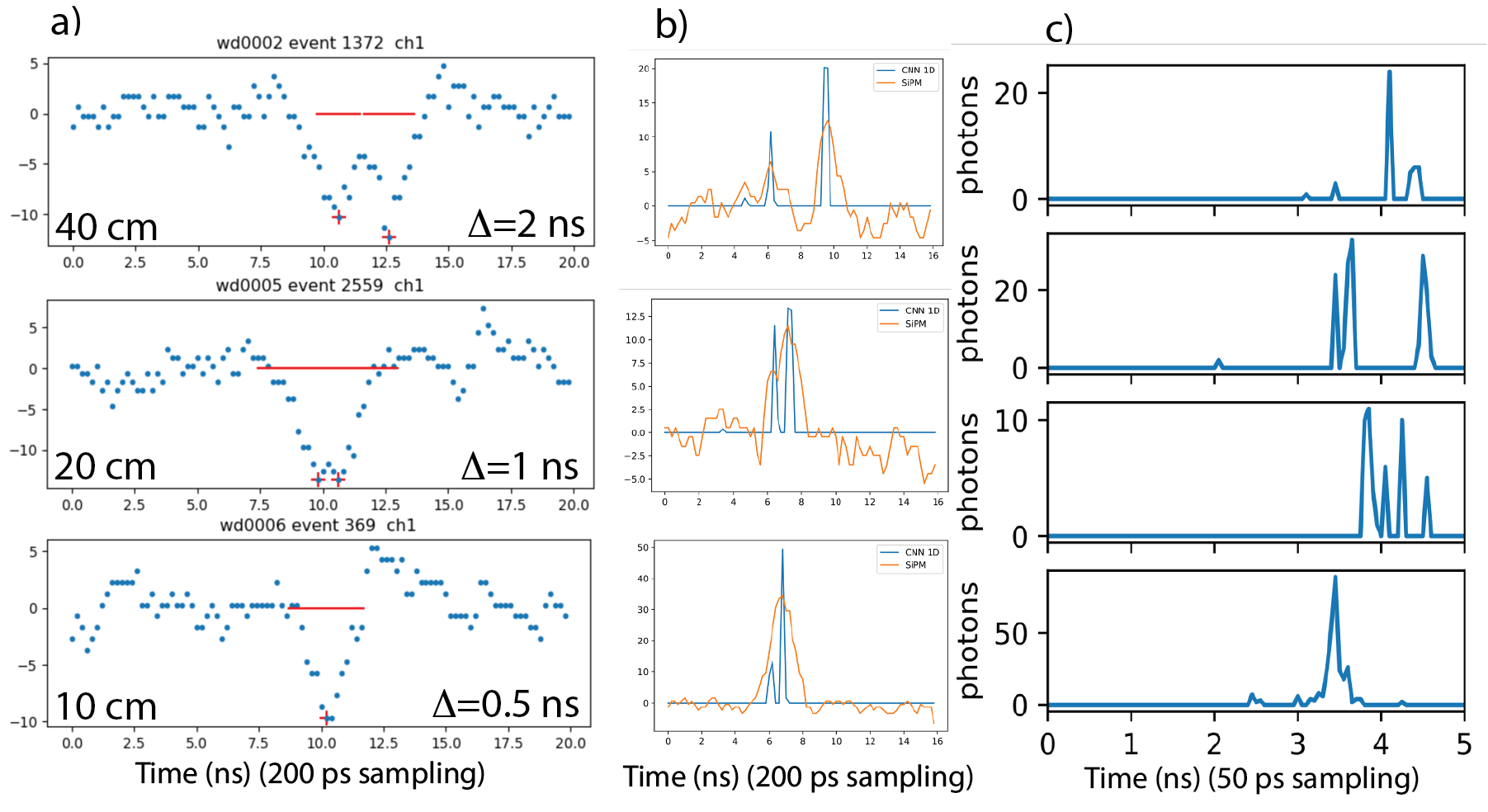}
\caption{\small (a) the pulse doublets were generated by an electron beam on fused-silica fibers that were of different lengths by 40, 20, and 10 cm, and read out by the same SiPM (MicroFC 3$\times$3 mm$^2$ with Mini-Circuit TB-409-S66+ amplifier), and digitized by CAEN V1742 (1 photon $\approx$ 10 ADC counts).  The speed of light in fuse-silica fibers is approximately 20 cm/ns.  It is clear that a 20-40 cm separation between energy deposits can be identified.   (b) When the energy deposition is separated by 10 cm or less, however, the identification becomes possible when a CNN algorithm is applied.  (c) Pulse trains for four 50 GeV pion events (GEANT4) in the HG-DREAM module are shown as example. }
\label{fig:CALOR2024pulses}       
\end{figure}

In order to evaluate the response of the fast output of OnSemi SiPMs to a few Cherenkov photons, we performed dedicated tests at the CERN PS T9 beam line in 2023.  Figure~\ref{fig:CALOR2024pulses} shows pulse trains from a 3$\times$3 mm$^2$ OnSemi SiPM.  These pulse trains were generated by two different-length fibers (differing by 40, 20, and 10 cm from the particle impact point on the fiber to the SiPM) and read out by a single SiPM.   It is evident that when events take place 40 or 20 cm apart, it is straightforward to identify the location of the energy deposit uniquely.  When the event separation is 10 cm or less, the identification becomes possible only when a neural network algorithm is applied, CNN in this case. Figure~\ref{fig:CALOR2024pulses}.c represents simulated (GEANT4) pulse trains generated by pions in HG-DREAM as an example of what is expected.

\begin{figure}[!htbp]
\centering
\includegraphics[width=8cm,clip]{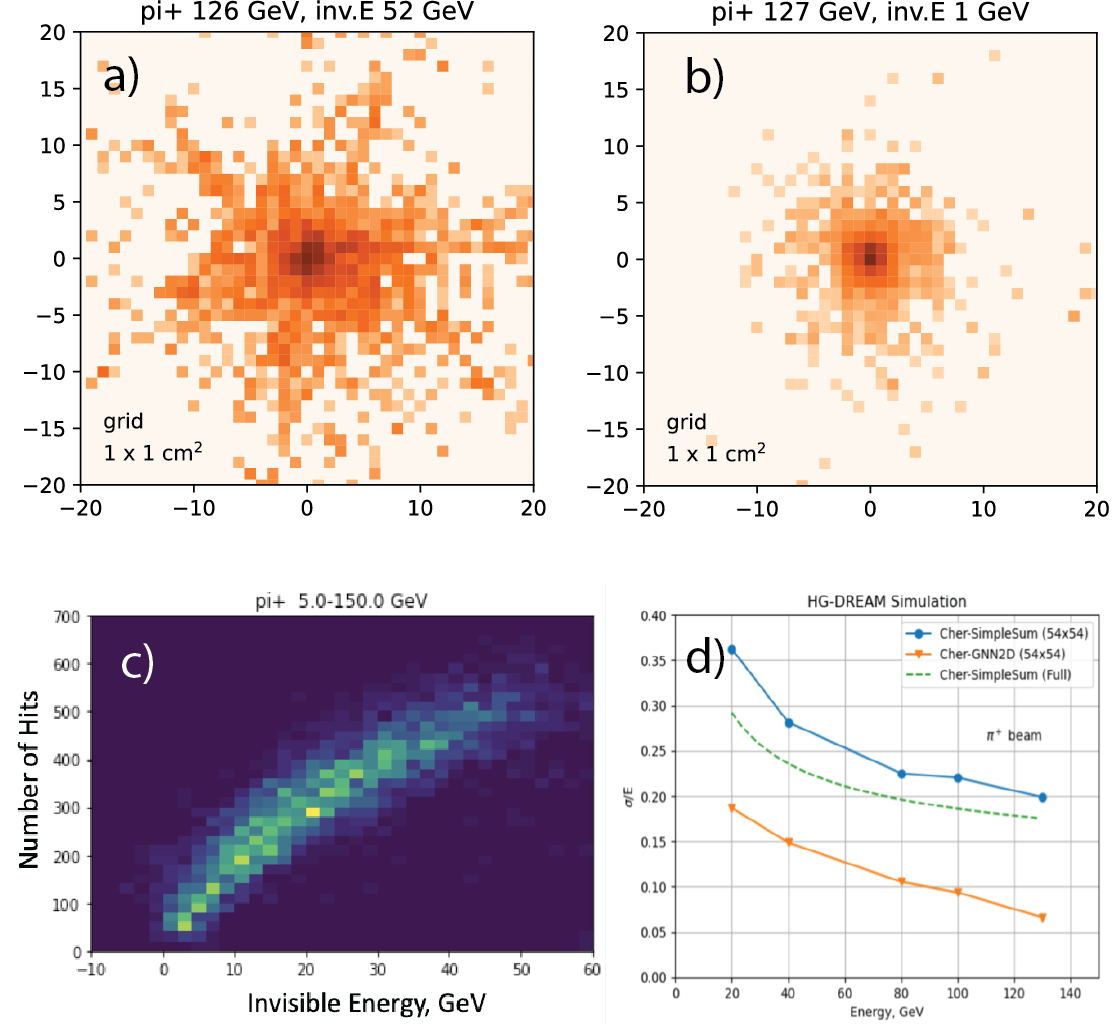}
\caption{\small Simulation of a 1$\times$1 cm$^2$ transversely segmented Cherenkov calorimeter:  a) the transverse image (2D) of a shower initiated by a 126 GeV pion where 52 GeV was undetected (invisible energy) exhibits a relatively active and large ``shower'' shape.  Note that each pixel corresponds to a 1$\times$1 cm$^2$ tower and that the $z$-axis is logarithmic.   If the invisible energy is small, as shown in panel b), and the shower shape is relatively compact.   c) The correlation between shower shape or number of hits (cells above a threshold) and the invisible energy is strong~\cite{instruments6040043}.  When a GNN algorithm is applied to a finely transversely segmented calorimeter (HG-DREAM without longitudinal segmentation), the energy resolution improves significantly as indicated by orange triangles in d).}
\label{fig:invisible_energy}       
\end{figure}

\begin{figure}[!htbp]
\centering
\includegraphics[width=8cm,clip]{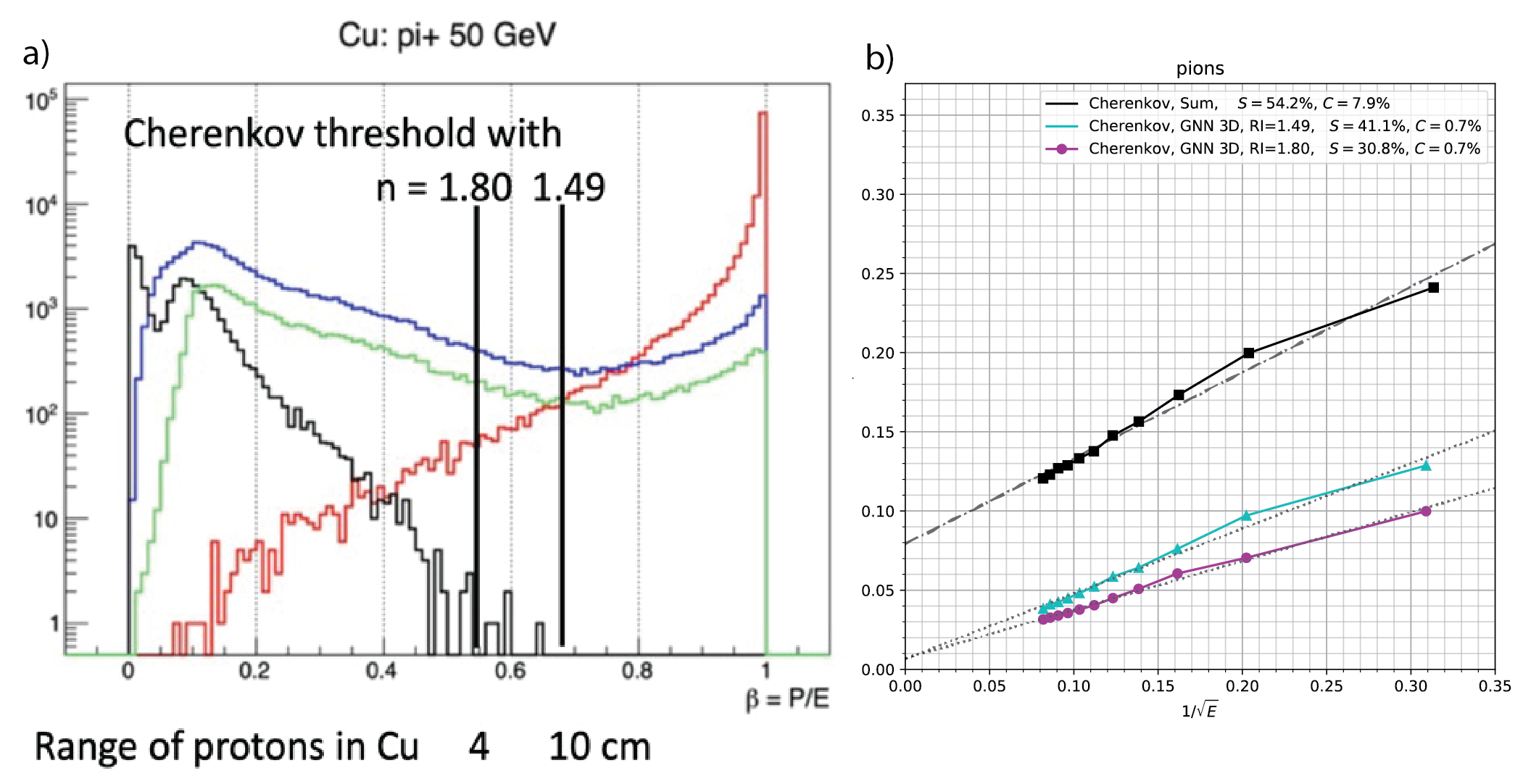}
\caption{\small (a) 50 GeV pions in copper generate different types of particles: protons (green), mesons (red), baryons (blue), and nuclear fragments (black).  If the refractive index (RI) increases from 1.49 to 1.8, the Cherenkov threshold decreases, and the calorimeter samples a larger number of protons, thereby enhancing the measured signal.  (b) Correspondingly, the energy resolution improves.  }
\label{fig:sapphire}       
\end{figure}

\begin{figure}[ht]
\centering
\includegraphics[width=8.0cm,clip]{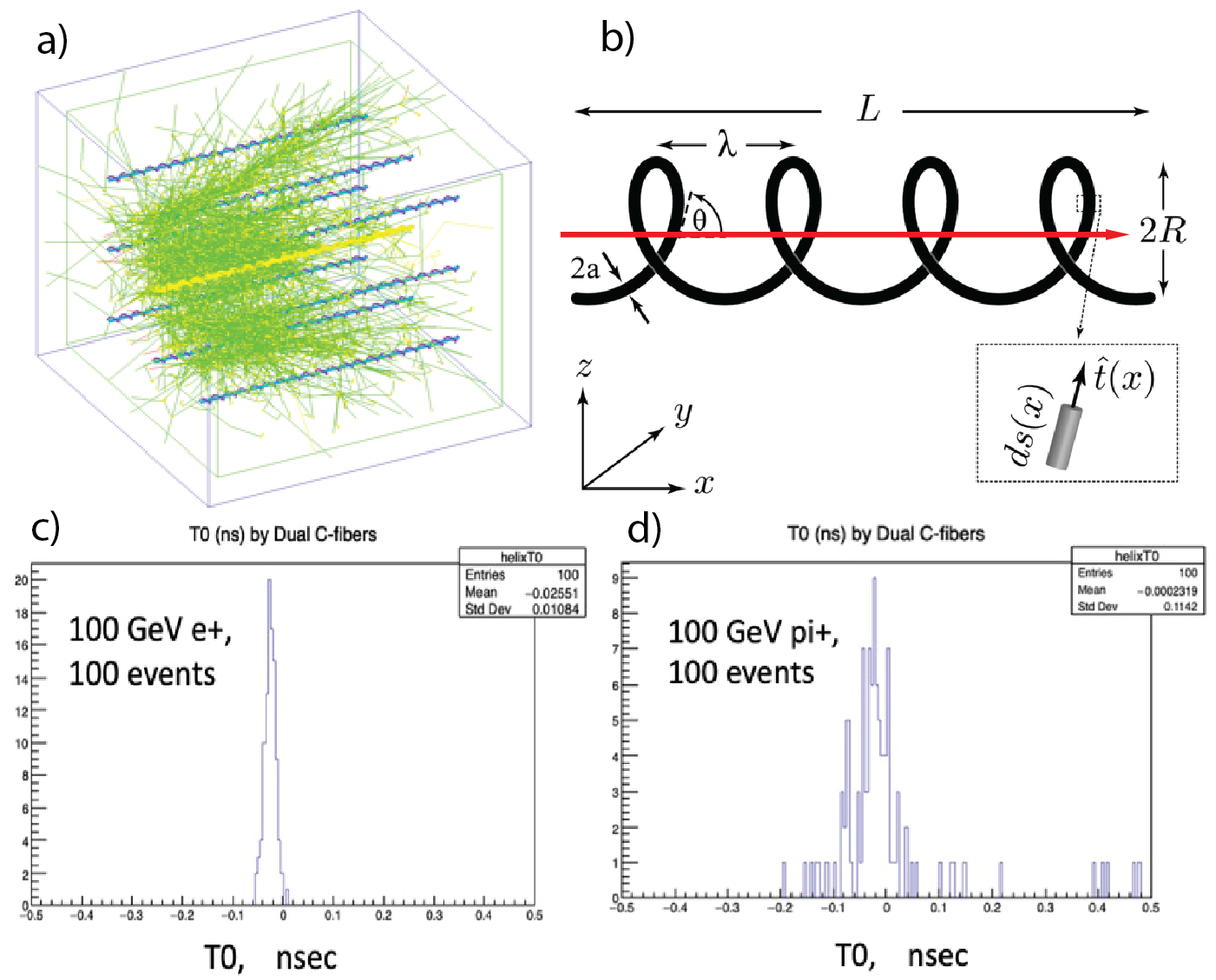}
\caption{\small The event display (a) of a simulated shower in copper and light propagation in embedded helical fibers. (b) The parameters for the helical fibers: $R=1.8$ mm, $\lambda=6.0$ mm, and $a=225\, \mu$m. (c) The $\sigma_{\rm rms}$ of arrival times difference between the helical and straight fibers for 100 GeV electron-initiated showers is about 11 ps, and (d) 114 ps for 100 GeV pion-initiated showers~\cite{Kunori_CALOR2024}.}
\label{fig:helical}       
\end{figure}

In addition to longitudinal segmentation using timing with HG-DREAM, we plan to evaluate the following performance enhancements that require precise timing information: the impact of timing on the energy resolution when timing information is used as an independent quantity in neural network algorithms in conjunction with high-granularity shower images~\cite{Akchurin_2021}, the assessment of the invisible energy using short integration times to access the fast proton signal as stand-in for slow neutrons in Cherenkov calorimetry (Figure~\ref{fig:invisible_energy}), and the effectiveness of absolute timing information from the calorimeter alone for pileup mitigation, TOF/PID measurements, jet substructure analyses by using different refractive index (Figure~\ref{fig:sapphire}), and/or helical fibers (Figure~\ref{fig:helical}).  

Figure~\ref{fig:sapphire} displays the impact of different refractive indices on the quality of energy measurement.  If fused-silica core fibers ($n \approx$1.49) are replaced with sapphire core fibers ($n\approx$1.8), the Cherenkov threshold for electrons decreases from 190 keV to 108 keV.  More significantly, the threshold for protons reduces from 350 MeV to 199 MeV, increasing the sensitivity to more abundant protons.  Figure~\ref{fig:sapphire}.b shows the impact of this change on hadronic energy resolution.  The stochastic term is reduced to $30.8\%$ (purple circles) from $41.1\%$ (green triangles).  It is also noteworthy to observe that a 3D GNN regression results in a sizable improvement over the customary (summation) reconstruction approach.

Figure~\ref{fig:helical} summarizes how helical fibers along with straight fibers may provide absolute timing/position information using the calorimeter alone.  The pitch of the helical fiber can be optimized not only to capture additional Cherenkov photons ({\it i.e.} $\theta\approx\theta_{\rm Ch}$) but also to introduce a delay compared to light propagation in a straight fiber.  The time difference between the two signals is then a direct measure of the $x$-position of the energy deposition inside the calorimeter.  The precision with which $t_0$ measurement can be achieved is shown in Figure~\ref{fig:helical}.c and Figure~\ref{fig:helical}.d, for electrons and pions, respectively.

\section{Conclusions}
We have built on the knowledge gained from the DREAM module by reconfiguring it for finer segmentation in (2+1)D with fast SiPMs and front-end readout.  In addition, we explore new optical fibers, not only for the enhancement of light yield but to sample different part of the particle spectrum in hadronic showers and to provide absolute timing measurement for the purposes of pileup mitigation, TOP/PID, jet substructure and other studies.  If successful, the HG-DREAM will prove to be a fertile test bed towards maximal information calorimetry.

\bibliography{ttu_bibs.bib}

\begin{thebibliography}{5}

\bibitem{Akchurin:2005eu}
N.~Akchurin et~al., {Electron detection with a dual-readout calorimeter}, Nucl.
  Instrum. Meth. \textbf{A536}, 29 (2005). \doiwoc{10.1016/j.nima.2004.06.178}

\bibitem{Akchurin:2005an}
N.~Akchurin et~al., {Hadron and jet detection with a dual-readout calorimeter},
  Nucl. Instrum. Meth. \textbf{A537}, 537 (2005).
  \doiwoc{10.1016/j.nima.2004.07.285}

\bibitem{instruments6040043}
N.~Akchurin, C.~Cowden, J.~Damgov, A.~Hussain, S.~Kunori, The (un)reasonable
  effectiveness of neural network in cherenkov calorimetry, Instruments
  \textbf{6} (2022). \doiwoc{10.3390/instruments6040043}

\bibitem{Akchurin_2021}
N.~Akchurin, C.~Cowden, J.~Damgov, A.~Hussain, S.~Kunori, On the use of neural
  networks for energy reconstruction in high-granularity calorimeters, Journal
  of Instrumentation \textbf{16}, P12036 (2021).
  \doiwoc{10.1088/1748-0221/16/12/P12036}

\bibitem{Kunori_CALOR2024}
S.~Kunori, {Vertex Imaging Calorimetry using AI/ML Tools} (2024), 20th
  International Conference on Calorimetry in Particle Physics (CALOR 2024),
  \urlstyle{tt}\url{https://indico.cern.ch/event/1339557/contributions/5898556/attachments/2860707/5004971/s_kunori_calore2024.pdf}

\end{thebibliography}

\let\clearpage\relax

\end{document}